\documentstyle[prb,aps]{revtex}
\begin{document}
\draft
\title{Glass phases of
flux lattices in layered superconductors}
\author{Anatoly Golub and Baruch Horovitz}
\address {Department of Physics, Ben-Gurion University
of the Negev, Beer-Sheva 84105, Israel}
\maketitle

\begin {abstract}
We study a flux lattice which is parallel to superconducting layers,
allowing for dislocations
and for disorder of both short wavelength and long wavelength.
We find that the long wavelength disorder of strength $\tilde{\Delta}$
has a significant effect on the phase 
diagram -- it produces a first order transition within the Bragg 
glass phase and leads to melting at strong $\tilde{\Delta}$. This then 
allows a Friedel scenario of 2D superconductivity.

\end{abstract}

The influence of 
disorder on the behavior of flux arrays
in superconductors is intensively 
studied in recent years. A model suggested
 by Larkin \cite{Larkin1}
in which random forces act independently on each vortex
was applied to  conventional superconductors.
Using the arguments of Ref. \cite{Larkin1} it was shown that the flux array 
is collectively pinned \cite{Larkin2}, forming a vortex glass (VG)
phase \cite{Nattermann,Feigelman,Fisher}. However,  experiments 
\cite{Cubitt,Forgan} on  weakly
disordered samples, reveal long-range order of the flux array, much
beyond the characteristic Larkin length. This phenomena is accounted 
for \cite{Korshunov1,Giamarchi1,Kierfeld} by an elastic dislocation-free 
theory of the flux lattice in 
weak random potential which predicts an algebraic decay of the
translational order and the existence of divergent Bragg peaks 
in the glass phase (Bragg glass). It was argued \cite{Giamarchi1,Kierfeld}
that the Bragg glass phase is stable against formation of dislocations 
in a finite range of the phase diagram.
  
The phenomena of melting of the flux lattice is of considerable interest. 
Melting has been observed \cite{Cubitt,Forgan,Safar,Kwok,Zeldov,Yeshurun}
as either a transition into a flux liquid phase or into a glass phase with 
a higher critical current. A model for melting, allowing for both disorder 
and dislocations \cite{Kierfeld,Carpentier} was recently studied. The 
model considers flux lines parallel to and confined between 
superconducting layers and allows for dislocations. This model was studied 
without disorder \cite{Mikheev} leading to flux melting at a critical 
temperature $T_c$ which is about factor $\sim 2$ from the solution of the 
more fundamental system in terms of superconducting phases 
\cite{Korshunov2,Horovitz1};
the latter allows also for flux loops and overhangs. Since disorder has
drastic effects on melting we expect that the model in terms of flux 
displacement is a reasonable starting point.

   The solution of Carpentier, Le-Doussal and Giamarchi (CLG) 
\cite{Carpentier} has shown explicitly that short wavelength disorder 
combined with dislocations leads to melting at a finite value of the 
disorder strength. CLG have used Replica Symmetry Breaking (RSB) methods 
as well as Renormalization Group (RG). They have also shown 
\cite{Carpentier,Giamarchi2} that this 
melting is compatible with a Lindemann criterion.

In the present work we allow for an additional term in the CGL model. This 
term is generated by RG within the CLG model and leads to significant 
effects on the phase diagram. Using RSB methods, we show that long 
wavelength disorder of strength $\tilde{\Delta}$ leads to a first order 
transition within the Bragg glass phase and as $\tilde{\Delta}$ increases 
it leads to melting. We find that the $\tilde{\Delta}$ induced melting is 
inconsistent with a universal Lindemann criterion. Finally we consider 
the quest for the Friedel scenario \cite{Friedel} in which a layered 
superconductor becomes a set of decoupled two-dimensional (2D) 
superconductors. This scenario fails in pure supercondutors 
\cite{Korshunov3,Horovitz2}, but is possible with some constraints in 
parallel fields \cite{Horovitz1} and in special models \cite{Dzierzawa}. 
With disorder which affects interlayer coupling the Friedel scenario 
becomes feasible in presence of a melted flux array.

The model \cite{Kierfeld,Carpentier,Mikheev} consists of layers with  
interlayer spacing $l$
where modulation in the  flux line density  couples 
to a random potential. We consider a Hamiltonian with two types of random 
potentials,

\begin{eqnarray}
H& = & \int d^{2}r \sum_{i}[\frac {c}{2}(\nabla \Phi_{i}({\bf r}))^{2}-
\eta_{i}({\bf r})
 \nabla \Phi_{i}({\bf r})  
     -\mu \cos (\Phi_{i}({\bf r})-\Phi_{i+1}({\bf r})) \nonumber \\
&  &- 2 Re(\zeta_{i}({\bf r}) e^{ i \Phi_{i}({\bf r})})]
 \end{eqnarray}
with Gaussian disorder correlations 
$<\zeta_{i}({\bf r})\zeta_{j}({\bf r}')>=
4Tg\delta_{i,j}\delta({\bf r}-{\bf r}') $ and $<\eta_{i}({\bf r})
\eta_{j}({\bf r}')>=
T \Delta\delta_{i,j} \delta({\bf r}-{\bf r}') $ where $T$ is the 
temperature. Here $\Phi_{i}({\bf r})$
stands for in-plane displacement of the vortex line in the $i$-th
layer, $c$ is an in-plane elastic constant, $g$ measures
disorder with Fourier component $\approx 2\pi/a$ where $a$ is the flux 
periodicity parallel to the layers, while 
$\Delta$ measures long wavelength disorder \cite{Giamarchi1}.
 The $\mu$ term is the coupling between layers which allows 
 for dislocations. In the pure system \cite{Mikheev} thermal fluctuations 
 lead to melting, i.e. $\mu$ is renormalized to zero, at $T_c =4\pi c$.

Dimensional Imry-Ma arguments are useful to check the stability
of an ordered phase which is a d dimensional elastic medium. In a domain of 
size $L$ the elastic energy is $\propto L^{d-2}$, the short wavelength disorder
(after averaging the square) is $\propto L^{d/2}$ while the long
wavelength disorder is $\propto L^{(d-2)/2}$. Thus short wavelength
disorder is relevant at $d<4$, while the long wavelength disorder
is marginal only at $d=2$, i.e. the latter is consistent with long range 
order in d=3.

To average over disorder we start with the replicated version of
Hamiltonian (Eq.(1)) which includes all relevant terms generated 
by renormalization,

%\newpage
\begin{eqnarray}
  H& = & \int d^{2}r\{ \frac {c}{2} \sum_{i,a}[(\nabla
\Phi_{i}^{a}({\bf r}))^{2}-\mu \cos (\Phi_{i}^{a}({\bf r})-
\Phi_{i+1}^{a}({\bf r}))]\nonumber \\ 
 & & \sum_{i,a,b}[\frac{\Delta}{2}\nabla\Phi_{i}^{a}({\bf r})
 \nabla\Phi_{i}^{b}({\bf r})     - \gamma\cos 
(\Phi_{i}^{a}({\bf r})-\Phi_{i+1}^{a}({\bf r})-\Phi_{i}^{b}({\bf r})
+\Phi_{i+1}^{b}({\bf r}))\nonumber \\
&  &-g\cos (\Phi_{i}^{a}({\bf r})-\Phi_{i}^{b}({\bf r}))] \}
\end{eqnarray}
where $a=1...n$ is the replica index.
Note in particular the $\gamma$ term which was not considered by GLC; this 
term is generated in second order RG from the $\mu$ term. Since it couples 
different replicas it can lead to RSB, i.e. this term leads to distinct 
phenomena and should be included in the full Hamiltonian.

We consider the variational free energy
$F_{var}=F_{0}+<H-H_{0}>$ with 
\begin {equation}
H_{0}=\frac{1}{2}\int\frac{d^{2}q}{(2\pi)^{2}}  
\int_{-\pi}^{\pi}\frac{dq_{z}}{2\pi}G_{ab}^{-1}({\bf q},q_{z})\Phi_{i}^{a}
({\bf q},q_{z})\Phi_{i}^{b}(-{\bf q},-q_{z}) 
\end{equation}
where the Greens' function $G_{ab}^{-1}({\bf q},q_{z})$ is determined by an 
extremum condition of $F_{var}$ and ${\bf q}, q_z$ are Fourier variables 
for ${\bf r}$ and $i$, respectively.

Defining the inverse Green's function in the form
$G_{ab}^{-1}({\bf q},q_{z})=\delta_{ab}G_{0}^{-1}({\bf q},q_{z})-
\sigma_{ab}-\Delta q^{2} $ with
$\sum_{a}\sigma_{ab}=0$ and $\sigma_{ab}=2(1-\cos 
q_{z})\sigma_{ab}^{\gamma}+\sigma_{ab}^{g}$ we obtain
the the self-consistent equations in the form
\begin{eqnarray}
G_{0}^{-1}(q,q_{z})& = & cq^{2}+2\tilde{\mu}(1-\cos q_{z}) \\
\sigma_{ab}^{g}& = &2g\exp[ -\frac{1}{2}B_{ab}^{g}]\\
\sigma_{ab}^{\gamma}& = &\gamma\exp[ -B_{ab}^{\gamma}]\\
\tilde{\mu}& =& \mu\exp [-\frac{1}{2}B_{aa}] 
\end{eqnarray}
where we define
\begin{eqnarray}
B_{ab}^{g}& =& 2T\sum_{{\bf q},q_{z}}[G_{aa}({\bf q},q_{z})-G_{ab}
({\bf q},q_{z})]\\
B_{ab}^{\gamma}& =& 2T\sum_{{\bf q},q_{z}}(1-\cos q_{z})[G_{aa}
({\bf q},q_{z}) -G_{ab}({\bf q},q_{z})]\\
B_{aa}& =& 2T\sum_{{\bf q},q_{z}}(1-\cos q_{z})G_{aa}({\bf q},q_{z})\,.
\end{eqnarray}
Here $\tilde{\mu}$ is the renormalized coupling between layers which is 
determined by the diagonal $B_{aa}$; $\tilde{\mu}=0$ signals a 2D phase, i.e. 
correlations in the $z$ direction are lost and the flux lattice has melted.

We study the full RSB solution of the saddle point equations (5-7). 
 The method of RSB \cite{Mezard} employs a representation of
hierarchical matrices such as $\sigma_{ab}^{g,\gamma}$ in term of 
functions $\sigma_{g,\gamma}(u)$, and similarly
 $B_{ab}^{g,\gamma}$ is represented by $B_{g,\gamma}(u)$ with $0<u<1$.
 
 We define two order parameters for RSB,
$[\sigma_{g,\gamma}](u)=u\sigma_{g,\gamma}(u)-
\int_{0}^{u}\sigma_{g,\gamma}(v)dv$. Using the inversion formula 
\cite{Mezard} for $G_{ab}$ and integrating over momenta we obtain

\begin{eqnarray}
\frac{1}{2} B_{g,\gamma}(u)& = & \frac{1}{u}g_{g,\gamma}(u)-
\int_{u}^{1}\frac{ds}{s^{2}}g_{g,\gamma}(s) \\
g_{g} (u) & = & -\tilde{T}\ln [\frac{m}{\Lambda}(\rho+1+w)] \nonumber\\
g_{\gamma}(u) &=& g_{g}(u)-\tilde{T}(\rho+1+w)^{-1}
\end{eqnarray}
Here $\rho(u) = [w(u)(w(u)+2)]^{1/2}$, $w(u)=( [\sigma_{g}](u))/ 2 m(u) $,
$m(u)=\tilde{\mu }+[\sigma_{\gamma}](u)$, $\Lambda$ is a cutoff of 
$cq^2$ ($\Lambda \gg m(u),\, [\sigma_{g}](u))$ and $\tilde{T}=T/T_c$ with 
$T_c=4\pi c$.

By differentiating Eqs.(5-6) we obtain
two coupled differential equations for the RSB functions $m(u)$ and $w(u)$
\begin{eqnarray}
\frac{2\tilde{T}}{u}\frac{dm}{du}& = & \frac{d}{du}[\frac{m \rho (1+w+\rho)}
{\rho(1+w+\rho)+Q(w+\rho)}] \nonumber \\
\frac{\tilde{T}}{u}\frac{dm}{du}(Q+w)& = & \frac{d}{du}[\frac{m \rho (Q+w)}
{Q+\rho}] 
\end{eqnarray}
where $Q(u)=m(dw/du)/(dm/du)$.

A general solution of these equations is rather difficult, so at first we 
consider special limits. When $\gamma =0$ we recover the CLG 
solution\cite{Carpentier}. Within the 3D Bragg glass phase $<[\Phi _i(r)
-\Phi _i(0)]^2>\sim \ln r$ so that positional correlations decay 
algebraically and long range order is weakly destroyed. The Bragg 
glass phase undergoes a continuous melting transition (for 
$\Delta=0$) at $g/\mu=2/e\tilde{T}$ as shown in the $\tilde{\Delta}=0$ 
plane of Fig. 1; for $\Delta \neq 0$ the transition becomes first order. 
Thus, the Bragg glass phase, due to both disorder and dislocations, melts 
into a 2D phase with $\tilde{\mu}=0$.

Consider next the case $g=0$, hence $w(u)=0$; the solution in this case is 
formally similar to that of a 2D disordered Josephson junction 
\cite{Horovitz3}. Eq. (13) yields then 
$(1-2\tilde{T}/u)m'(u)=0$, i.e. $m(u)$ is a one step function, with the 
step at $u=2\tilde{T}$. Since $u<1$ the onset of this solution is at 
$\tilde{T}=1/2$. i.e. at $T=T_c /2$. Eqs. (6,7) determine the jump in 
$[\sigma_\gamma](u)$ from zero ($u<2\tilde{T}$) to a value 
$\sigma_\gamma^0$ at $2\tilde{T}<u<1$, where

\begin{eqnarray}
\frac{\tilde{\mu}+\sigma_\gamma ^0}{\Lambda}&=&e^{-1}\left(4e\tilde{T}
\frac{\gamma}{\Lambda}\right)^{1/(1-2\tilde{T})}  \\
\frac{\tilde{\mu}}{\Lambda}&=&e^{-1}\left[e^{\tilde{\Delta}+1/2}
\left(4e\tilde{T}\frac{\gamma}{\Lambda}\right)^{-1/2}\,\frac{\mu}
{\Lambda}\right]^{1/(1/2-\tilde{\Delta})}
\end{eqnarray} 
where $\tilde{\Delta}=4\pi \tilde{T}^2\Delta$. This solution is valid 
for $\tilde{\Delta}<1/2$ and $\tilde{T}<1/2$. For $\gamma$ of order $\mu$, 
near the $\tilde{T}=1/2$ transition $\tilde{\mu}$ is finite while 
$\tilde{\mu}+\sigma_\gamma^0$ vanishes. Thus $\sigma_\gamma^0 <0$ is finite 
up to $\tilde{T}=1/2$ and vanishes at $\tilde{T}>1/2$, i.e. the transition 
is of first order. When $\tilde{\Delta}>\tilde{T}$ within this phase 
$\sigma_\gamma^0$ changes sign and becomes positive.

The phase at $\tilde{\Delta}<1/2$ and $\tilde{T}<1/2$ is a coexistence 
phase--it has both long range order ($\tilde{\mu} \neq 0$) and glass 
order ($\sigma_\gamma^0 \neq 0$). (As noted above this is consistent with 
the Imry-Ma argument). At $\tilde{\Delta}=1/2$ we find a disorder 
driven transition where $\tilde{\mu}$ vanishes 
continuously, leading to a 2D glass phase at $\tilde{\Delta}>1/2$.

We note also that a replica symmetric solution is possible for 
$1/2<\tilde{T}<1$ with
\begin {equation}
\frac{\tilde{\mu}}{\Lambda}=e^{-1}\left(e^{\tilde{\Delta}+1/2}
\frac{\mu}{\Lambda}\right)^{1/(1-\tilde{T}-\tilde{\Delta})}
\end{equation}
i.e $\tilde{\mu} \neq 0$ for $\tilde{T}+\tilde{\Delta}<1,\,\tilde{T}>1/2$ 
as shown in Fig. 1. Comparison with Eqs. (14-15) shows that 
$\tilde{\mu}$ is also discontinuous at the $\tilde{T}=1/2$ transition.

Finally we consider the case where both $g$ and $\Delta$ are finite. We 
can demonstrate the existence of a first order transition at small $g$ by showing 
a coexistence of two solutions. The first solution is an expansion near the 
$g=0$ solution, i.e. $w(u)\ll 1$ with 
$[\sigma_g]=O(g),\,[\sigma_\gamma]=O(1)$; for $u<\tilde{T}/2$ the solution 
for $[\sigma_g]$ is similar to the $\gamma=0$ case, i.e. 
$[\sigma_g](u) \sim 
u^2$ for small $u$, consistent with Bragg glass correlations.
This solution is valid (assuming $\mu \sim \gamma$) if
\begin{equation}
g/\Lambda \ll (\gamma/\Lambda)^\frac{1-3\tilde{\Delta} +2\tilde{T}
\tilde{\Delta}}{(1-2\tilde{T})(1-2\tilde{\Delta})} \nonumber\\
\end{equation}
i.e. for weak coupling $g/\Lambda ,\gamma/\Lambda \ll 1$ this expansion 
breaks down close to the transitions at $\tilde{T}=1/2$ and 
$\tilde{\Delta}=1/2$. The second solution is an expansion around the 
$\gamma=0$ solution with $[\sigma_{\gamma}](u) \ll \tilde{\mu}$. This leads 
to $[\sigma_g]=O(g^2),\,[\sigma_{\gamma}]=O(g)$ and is valid for 
$\tilde{\Delta}<\tilde{T}$. Thus for small $g$ there is a two solution 
regime which implies a first order transition at some $\tilde{T}\lesssim 1/2$. 
We indicate this transition by a spaced dashed line in Fig. 1, though we 
do not know its precise location.

As shown in Fig. 1, we find that the main feature of the CLG scenario is 
valid-- for small disorder the Bragg glass is stable, while at large 
disorder, which can have either short or long wavelength, dislocations are 
enhanced by disorder and lead to melting.

These analytic results for melting allow us to test the Lindemann criterion, 
which is of common use \cite{Blatter}. For the $\gamma=0$ case, CLG 
consider a Lindemann criterion of the form \cite{Carpentier,Giamarchi2} 
$<[\Phi_{i+1}({\bf r})-\Phi_i({\bf r})]^2>=c_L^2$, 
with average done in the elastic 
limit, i.e. the cosine of the $\mu$ term in Eq. (1) is expanded, This 
criterion leads \cite{Carpentier} to a reasonable value of $c_L\lesssim 
1$. For the $g=0$ case an elastic limit leads to an expansion of both the 
$\mu$ and $\gamma$ terms in Eq. (2) so that RSB is not induced. Since long 
range order is present the Lindemann criterion is $<\Phi_i^2(r)>=c_{L}^2$; 
however the replica symmetric solution yields $<\Phi_i^2(r)>=
(\tilde{\Delta}+\tilde{T})\ln (\Lambda/\mu)$, i.e. at melting 
$c_L^2\approx \ln (\Lambda/\mu)$. since $\Lambda/\mu$ depends on the 
anisotropy of the system the Lindemann number $c_L$ is non-universal.

In order to relate the phase diagram to the actual magnetic field B we 
need to identify $c$ by the elastic constants \cite{Blatter,Sudbo} which 
are dispersive, $c_{44}^{\|}\approx 
c_{11}^{\|}=(B^2/4\pi)/(1+\lambda_c^2q^2+\lambda^2q_z^2)$; here 
$\lambda,\,\lambda_c$ are penetration lengths and 
$\epsilon=\lambda/\lambda_c <1$ is the anisotropy ($c_{44}^{\|}$ has a 
smaller second term which is neglected here). The behavior near melting is 
dominated by $q\rightarrow 0$ and $q_z\approx 1/l$. The lattice 
periodicities satisfy $l=a\epsilon$ if $l>d$ for weak fields, i.e. $B=\phi_0 
/a^2\epsilon <\phi_o\epsilon /d^2$ ($d$ is the  
spacing of the superconducting layers, which is the lower bound on the 
interlayer spacing $l$ of the flux lattice, and $\phi_0$ is the flux quantum),
or $l=d$ for strong fields, 
$B=\phi_0/ad >\phi_0\epsilon/d^2 $. 
By rescaling the $x,z$ coordinates we identify

\begin{eqnarray}
4\pi c &\approx &a\phi_0^2/(4\pi ^2\lambda \lambda_c) \sim B^{-1/2}
\mbox{\hspace{26mm}}B<\phi_0\epsilon /d^2      \nonumber\\
4\pi c &\approx & d\phi_0^2/(4\pi ^2\lambda ^2) 
\mbox{\hspace{42mm}}B>\phi_0\epsilon /d^2 \, .
\end{eqnarray}
$T_c$ ($=4\pi c$) is smallest for large fields, but even then its value 
is $<100 K$ for typical $CuO_2$ superconductors only very near the 
superconducting transition where $\lambda$ diverges. Thus it is the 
disorder induced melting which is relevant to experimental data.

Since $\Delta$ couples to $\nabla \Phi_i(r)$ \cite{Giamarchi1} it is $B$ 
independent so that $\tilde{\Delta}=\Delta /4\pi c^2 \sim B$ for small 
fields and $\tilde{\Delta} \sim $constant for strong fields. Thus the 
$\Delta$ induced melting can be induced by increasing the magnetic field 
if $\tilde{\Delta} =1/2$ is achieved for weak fields. On the other hand, 
for the $g$ induced melting $g\sim 1/a^2$ \cite{Giamarchi1}, and by using 
the $c_{66}$ elastic constant \cite{Kogan}  we identify $\mu =a^2\phi_0 
B\epsilon^3/[4\pi ^2 l (8\pi \lambda)^2]$. For weak fields the melting 
temperature is $T_m^{-1} =eg/(8\pi c\mu) \sim B^{3/2}$ while for strong 
fields $T_m^{-1} \sim B^3$. In this case melting is induced by increasing 
$B$ for both weak or strong fields.

We note finally that any disorder which melts a flux lattice parallel to 
superconducting layers is a route to the Friedel scenario \cite{Friedel}. 
In this scenario the layers remain superconducting so that the system 
behaves as a 2D superconductor. This scenario is not valid in  pure layered 
superconductor \cite{Korshunov3,Horovitz2}, it is valid for parallel 
fields with restricted system parameters (e.g. large vortex core energy) 
\cite{Horovitz1} or in special models \cite{Dzierzawa}. In the present study, disorder which affects interlayer 
coupling leads to melting which decouples the layers; assuming weak 
intralayer disorder the layers remain superconducting. Thus we have a 
model system which exhibits the Friedel scenario.

In conclusion we have shown that a new interaction  term, generated by RG, 
leads to a significant role of the long wavelength disorder.
This interaction extends the CLG results to the more complex phase diagram 
of Fig.1. We find that the Bragg glass is stable for weak disorder of 
either short or long wavelength. The long wavelength disorder induces a 
first order transition within the Bragg glass phase; it also leads to 
melting which is inconsistent with a Lindemann criterion. We propose that 
experiments with parallel magnetic fields can test the present theory of 
melting as well as test the possibility of 2D superconductivity.

\vspace{10 mm}
Acknowledgments: We are grateful  
to E. Zeldov, Y. Y. Goldschmit, A. Kapitulnik, M. V. Feigel'man and
D. R. Nelson for useful discussions and to T. Giamarchi, T. Nattermann, P. Le
Doussal and V. M. Vinokur for discussions of their works and 
for illuminating comments. This research was supported by a grant from the 
Israel Science Foundation.

\newpage

                Figure Caption

Phase diagram of layered flux lattices with disorder of short 
wavelength (with strength $g$) and of long wavelength (with strength
$\tilde{\Delta}$). The 
dashed line  is the line of first order transition (discontinuity
of the interlayer coupling and of the glass order); the spaced 
dashed line is its approximate extension to $g \neq 0$. The 3D glass 
regime at $g=0$ has long range order while the 3D Bragg glass at 
$g\neq 0$ has algebraically decaying positional order.

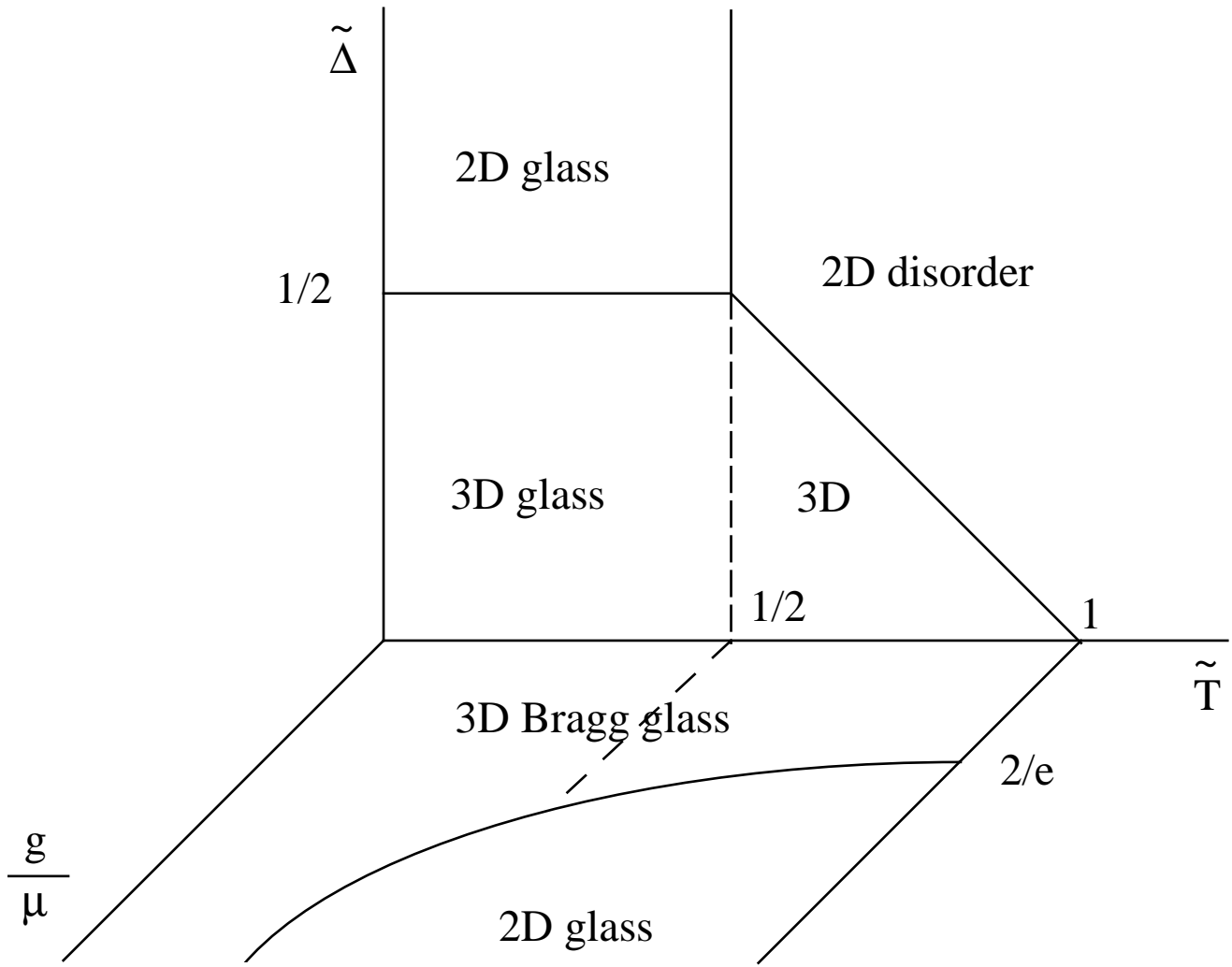
\end{document}